\documentclass [12pt,a4paper,draft]{article}
\usepackage{times}

\DeclareFontFamily{OT1}{times}{}
\DeclareFontShape {OT1}{times}{m }{n }{ <-> ptmr }{}
\DeclareFontShape {OT1}{times}{bx}{n }{ <-> ptmb }{}
\DeclareFontShape {OT1}{times}{m }{it}{ <-> ptmri}{}
\DeclareFontShape {OT1}{times}{bx}{it}{ <-> ptmbi}{}

\begin{document}

\title{{\bf Comment on ``Deuterium--tritium fusion reactors without external fusion breeding'' by Eliezer et al.}}
 
\author{{A. Gsponer and J.-P. Hurni}\\ \emph{Independent Scientific Research Institute}\\ \emph{ Box 30, CH-1211 Geneva 12}\\ 
e-mail: isri@vtx.ch}

\date{July 30, 1998 [Rev. August 18, 1998]}

\maketitle

\rule{124mm}{0.2mm}

\begin{abstract}

\noindent Inclusion of inverse Compton effects in the calculation of deuterium-deuteri\-um burn under the extreme conditions considered by Eliezer et al. [Phys. Lett. A 243 (1998) 298] are shown to decrease the maximum burn temperature from about 300 keV to only 100--150 keV. This decrease is such that tritium breeding by the DD~$\rightarrow$ T + p reaction is not sufficient to replace the small amount of tritium that is initially added to the deuterium plasma in order to trigger ignition at less than 10 keV.

\end{abstract}

\rule{124mm}{0.2mm}

\vspace{2 cm}
\begin{center}
Published in ~ {\bf Phys. Lett. A, Vol. 253} (1999) 119-121
\end{center}
\vspace{2 cm}


Simplified models of thermonuclear burn are well known to lead to first order results that are generally in reasonable agreement with more sophisticated models.  However, the prerequisite for such agreement is that, in the parameter range under consideration, all physical phenomena that are relevant to this order of approximation be consistently taken into account.

In a recent Letter, the results of space-time independent calculations of very-high temperature ($T_i \approx 300$ keV, $T_e \approx 150$ keV) thermonuclear burn of a very-high density (compression of $\approx$ 30'000 times solid density) DD plasma containing a small initial addition of tritium are presented \cite{ELIEZ98}. From these calculations, the authors conclude that the initial amount of about 2\% of tritium needed to trigger ignition at less that 10 keV is regenerated by the DD reaction if the burning temperature is higher that 200 kev.  If this is correct, it would be possible to conceive DT fusion reactors which would not need external tritium breeding for their operation.

However, in these calculations, a number of effects and processes which become important at extremely high temperatures and densities seem not to have been fully taken into account. These phenomena comprise various relativistic and degeneracy effects, as well as several radiation processes which include inverse bremsstrahlung, electron-electron bremsstrahlung, Compton scattering, etc. Of these phenomena, the most important is electron energy loss by inverse Compton scattering.  The purpose of this Comment is to show that its inclusion in the simulation totally changes the main conclusion of Ref.\ \cite{ELIEZ98}.

The energy lost by electrons of temperature $kT_e$ in a photon gas by the inverse Compton effect is

\begin{equation}
W_C \equiv \frac{dE_C}{dt} = \frac{32}{3} \pi{r_e}^2 c \frac{E_r}{mc^2} N_e  (kT_e-kT_r)  ~ ,
\end{equation}

\noindent where $E_r=\frac{4}{c}\sigma{T_r}^4$ is the radiation energy density and $N_e$ the electron number density \cite{WEYMA65}. An estimate of $E_r$ can be made by assuming that the average traveling time of the bremsstrahlung photons through a thin plasma of radius $R$ is $\Delta t \approx R/c$. Therefore, if $W_B$ is the bremsstrahlung radiation loss rate, $E_r \approx{W_B}R/c$. Using this estimate, it is readily seen that $W_C > W_B$ for $T_e > 60$ keV in the parameter range considered in Ref.\ \cite{ELIEZ98}, so that Compton effects cannot be neglected.

To make this statement more quantitative, and in order to calculate how much of the initial tritium can effectively be regenerated during thermonuclear DD burn, we have used our simulation program ISRINEX \cite{GSPON97} to reproduce the results of Ref.\ 1 and to show the consequences of taking radiation effects more fully into account.

In our program, inverse bremsstrahlung and Compton effects are simulated by a model attributed to H. Hurwitz.  This model was first published in \cite{FRALE74} and is used in both small (e.g., \cite{KIRKP79}) and large (e.g., \cite{TAHIR86}) simulation programs.  However, since there is some discrepancy in the published literature on the details of this model, we take this Comment as an opportunity to review it briefly.

Hurwitz's model is a three-temperature model in which the electron and photon energy equations are coupled as follows:


\begin{eqnarray}
 \frac{dE_e}{dt} & = & - \frac{dE_B}{dt}  - \frac{dE_C}{dt} ~ ,\\
 \frac{dE_r}{dt} & = & + \frac{dE_B}{dt}  + \frac{dE_C}{dt} - R_{loss} ~ .  \end{eqnarray}

The first term corresponds to bremsstrahlung effects. It is obtained by integrating over frequency the difference between emission by the electrons of bremsstrahlung radiation and reabsorption of the local radiation field (inverse bremsstrahlung). With the notation of Ref.\ \cite{ZELDO66}, this is
 
\begin{equation}
\frac{dE_B}{dt} = \int_{0}^{\infty} c\kappa_\nu \, (U_{\nu p} - U_{\nu})\, d\nu = W_B \, G(\gamma) \, \frac{T_e-T_r}{T_e} ~ ,
\end{equation}

\noindent
where $\gamma=T_r/T_e$ and

\begin{equation}
G(\gamma) = \frac{1}{1-\gamma} \int_{0}^{\infty} \frac{1}{2} \exp(x/2)~ {\rm K}_0(x/2) ~\frac {1-\exp{ (x/{\gamma}-x)} } {1-\exp(x/{\gamma})}~dx ~ ,  \end{equation}

\noindent where ${\rm K}_0$ is the Bessel function characteristic of the quantum-mechanical brems\-str\-ahlung emission spectrum. For $\gamma \rightarrow 0$, $G(\gamma) \rightarrow 1$, making  $dE_B/dt$ equal to the pure bremsstrahlung rate $W_B$; for $\gamma \rightarrow \infty$, $G(\gamma) \rightarrow \pi^2/4$, giving pure inverse bremsstrahlung.  Except for integer values of $\gamma$, $G(\gamma)$ cannot be calculated analytically \cite{HURNI98}. A simplified fit, $G(\gamma) \approx {\pi^2}/{4}~(\gamma+1.03)/(\gamma+2.54)$ is therefore used in the calculations.

The second term in (2,3) corresponds to Compton effects as given by (1).  The third term in (3) is the photon energy loss rate which for the purpose of this Comment is approximated by:

\begin{equation}
R_{loss} = \frac{3}{R} \sigma{T_r}^4 ~ ,
\end{equation}

\noindent or

\begin{equation}
R_{loss} = \min \left[ \frac{3}{R} \sigma{T_r}^4 ~ , W_B \right] ~.
\end{equation}

\noindent Expression (6) is the maximum possible photon energy loss, so that its use tends to underestimate the value of $T_r$ and thus the effect of Compton scattering. On the other hand, expression (7), which gives the correct limit in the case of an infinitely thin plasma (and which gives results in close agreement with simulations presented in \cite{FRALE74}), tends to minimize radiation losses and therefore to overestimate Compton effects. 

The results of our calculations for the simulation shown in Figs.\ 2--4 of Ref.\ \cite{ELIEZ98} are given in Table 1. The first column gives our results when the only radiation effect included is bremsstrahlung losses, i.e., $dE_e/dt = W_B$, as in Ref. \cite{ELIEZ98}. They are in reasonable agreement with Ref.\ \cite{ELIEZ98}, except possibly for the final tritium content which is only 0.83 instead of slightly more than 1.  This may partly be explained by the fact that in our calculation the maximum ion temperature is only 270 keV instead of about 300 keV. Otherwise, the ratio of tritium content goes through a minimum at a burn time of about 2 ps, and through a maximum at about 6 ps, as in Fig.\ 2 of Ref.\ \cite{ELIEZ98}.

The second column corresponds to Hurwitz's three-temperature model with inverse bremsstrahlung and radiation losses included according to (4) and (6). The effects on $T_i$ and $T_e$ are small, and the final fuel burn-up fraction is unchanged. However, the maximum radiation temperature, i.e., 17 keV, is large enough to imply that Compton effects cannot be neglected.

The third column gives our results when both bremsstrahlung and Compton effects are fully included. The maximum ion temperature is only half the $\approx$ 300 keV found in \cite{ELIEZ98}. Consequently, $T_i$ is not large enough to breed enough tritium to replace the initial tritium content. The final tritium content is only about 1/4 of its initial value.

Finally, the last column shows what happens when the photon losses are reduced, e.g., when the thermonuclear fuel is surrounded by a some material opaque to low
energy photons but relatively transparent to bremsstrahlung. $T_i$ is further reduced to $\approx$ 100 keV, the final tritium content ratio is less than 1/5, and the final burn-up ratio is only 0.34.

In conclusion, we have shown that by taking radiation effects such as inverse Compton scattering fully into account the maximum burn temperature of a highly compressed DD pellet is reduced from about 300 keV to only 100 keV.  This decrease in ion temperature leads to a substantial reduction in the fuel burn-up ratio. Moreover, this decrease is such that tritium breeding by the DD reaction is not sufficient to replace the amount of tritium that is added to the deuterium plasma in order to trigger ignition at less than 10 keV. These conclusions would be reinforced by taking relativistic and degeneracy effects into account. Finally, if the size of the pellet were larger, or if the fuel were surrounded by various materials, as is the case in a realistic pellet design, the importance of radiation effects would be further enhanced. These problems illustrate the difficulty of burning advanced thermonuclear fuels, such as DT$_x$, D${^3}$He, or p$^{11}$B \cite{MARTI96}, relative to burning those containing a large tritium fraction, such as DT or ${^6}$LiDT.

%
%

%
%

%
\begin{table}
\begin{tabular}{|c|c|c|c|c|} 		
\hline
\multicolumn{5}{|c|}{\raisebox{+0.4em}{{\bf High-temperature, high-density, DD burn \rule{0mm}{6mm}}}} \\ 
\hline
    & No          & Direct and     & Direct and   & Reduced   \\ 
    & radiation   & inverse        & inverse      & photon    \\   
    & interaction & bremsstrahlung & Compton      & losses    \\
\hline
\multicolumn{5}{|l|}{\emph{Maximum temperatures:}\raisebox{+1.em}{~}\raisebox{-0.5em}{~}} \\  
\hline 

$T_i$  & 267 & 272 & 146 & 100  \\ 
$T_e$  & 145 & 142 & ~61 & ~48  \\
$T_r$  & --- & ~17 & ~23 & ~46  \\   
\hline
\multicolumn{5}{|l|}{\emph{Final fuel burn-up fraction:}\raisebox{+1.em}{~}\raisebox{-0.5em}{~}} \\ 
\hline
    & 0.47 & 0.47 & 0.39 & 0.34 \\
\hline
\multicolumn{5}{|l|}{\emph{Ratio of tritium content:}\raisebox{+1.em}{~}\raisebox{-0.5em}{~}} \\ 
\hline
Initial & 1.00 & 1.00 & 1.00 & 1.00 \\ 
Minimum & 0.47 & 0.50 & 0.50 & 0.50 \\   
Maximum & 1.47 & 1.50 & 0.87 & 0.57 \\
Final   & 0.83 & 0.83 & 0.27 & 0.18 \\
\hline
\end{tabular}
\caption{Maximum ion, electron, and radiation temperatures; final fuel burn-up fraction; and ratios of tritium content as a function of time for several radiation interaction models. The DT$_x$ pellet has a mass of 0.167 mg  and the parameters at maximum compression before ignition are as follows: $x$ = 0.02; $\rho R$ = 10 g/cm$^2$; $T_i=T_e$ = 10 keV; and $T_r$ = 1 keV.} 
\end{table}

\end{document}